# On the Positioning
# of Objects in Space

## Abstract

The personal spatial structure of an observer is introduced as a central element in the positioning of objects in space. The link between a reference frame used by an observer and his personal spatial structure is discussed. Research on inversion or reversal of incoming light in psychology indicates that the personal spatial structure of an individual, as well as a reference frame that he uses, depends on the internal coordination of sensory stimuli and that the position of objects in space depends in part on psychological factors that affect one's personal spatial structure. Other research in psychology demonstrating the flexibility for an observer in determining the direction of up-down relative to a particular figure, and indeed relative to the entire surround, is also noted and supports the thesis that the observer plays a role in the positioning of objects in space.

## Text

On a practical basis, an individual knows about space through the employment of a spatial structure delineated by a set of orthogonal spatial axes for height (up-down), width (right-left), and depth (in-out). Using this spatial structure, the individual gauges to a high degree of accuracy the spatial position of objects in the world, such that he is able to perform intricate coordinated sensori-motor actions in the world.[1] The use of a formal spatial coordinate system is an extension of one's personal spatial structure. It is because of one's personal spatial structure that an individual can employ and understand more formal spatial reference frames.[2]

Evidence supporting these theses comes from work on inversion and reversal of all incoming light (e.g., Dolezal 1982; Erismann & Kohler, 1953, 1958; Pronko & Snyder, 1951; Snyder & Pronko, 1952; Stratton, 1896, 1897a, 1897b). It was found that when there is inversion or reversal of light to the retina, the observer adapts both behaviorally and perceptually such that the observer can function largely, if not totally, in the way in which he was functioning before incoming light was inverted and without awareness that the pattern of incoming light has been altered. Before adaptation the position of objects in the visual field is altered, the degree depending on the nature of the alteration of incoming light. Adaptation can be accounted for in terms of a shift of the observer's personal spatial structure. A person's spatial structure depends on the internal coordination of sensory stimuli, for it is this coordination of sensory stimuli, notably touch and vision, that allows for adaptation. The shift in one's personal spatial structure upon adaptation then also affects *the formal spatial reference frame* used by these observers in their scientific investigation of the physical world. One such joining of the *x*, *y*, and *z* axes of a spatial reference frame and one's personal spatial structure is shown in Figure 1. In this figure:

1. The *z* axis is in the vertical direction relative to the subject, appearing to go up and down.

2. The *y* axis runs perpendicular to the ideal plane formed by the subject's face, appearing to go in and out.

3. The *x* axis runs horizontally relative to the subject, from side to side.

---

[1] Asch and Witkin (e.g., 1948a) investigated this spatial structure and called it a reference frame.

[2] In our daily life and in Newtonian mechanics, a reference frame is a spatial coordinate system associated with a physical object. A temporal coordinate system is associated with the reference frame. In relativity theory, a reference frame is a spatiotemporal coordinate system.

This *sample* spatial structure and the accompanying spatial coordinate system is really Euclidean in nature, where unit distance is maintained and where one can use measuring instruments of unit length to form a grid of squares throughout space.

In the verification of the general theory of relativity by Eddington and his colleagues, it is interesting that in the determination of the observation of the motion of light rays passing close to the sun during a solar eclipse, observers did not use the curved reference frame near the sun that results from the sun's gravitational field and within which the motion of light rays may appear straight. Instead, the light rays traced a curved path as they travelled by the sun for an observer on the earth (Einstein, 1917/1961). An observer on earth applied his essentially Euclidean (non-curved) spatial framework to the light rays traveling in curved spacetime near the sun and observed that the paths of the light rays near the sun curve. This observer's perception occurred because the spatial coordinate system used by an individual is rooted in the personal spatial structure of an individual, which in this case is Euclidean.

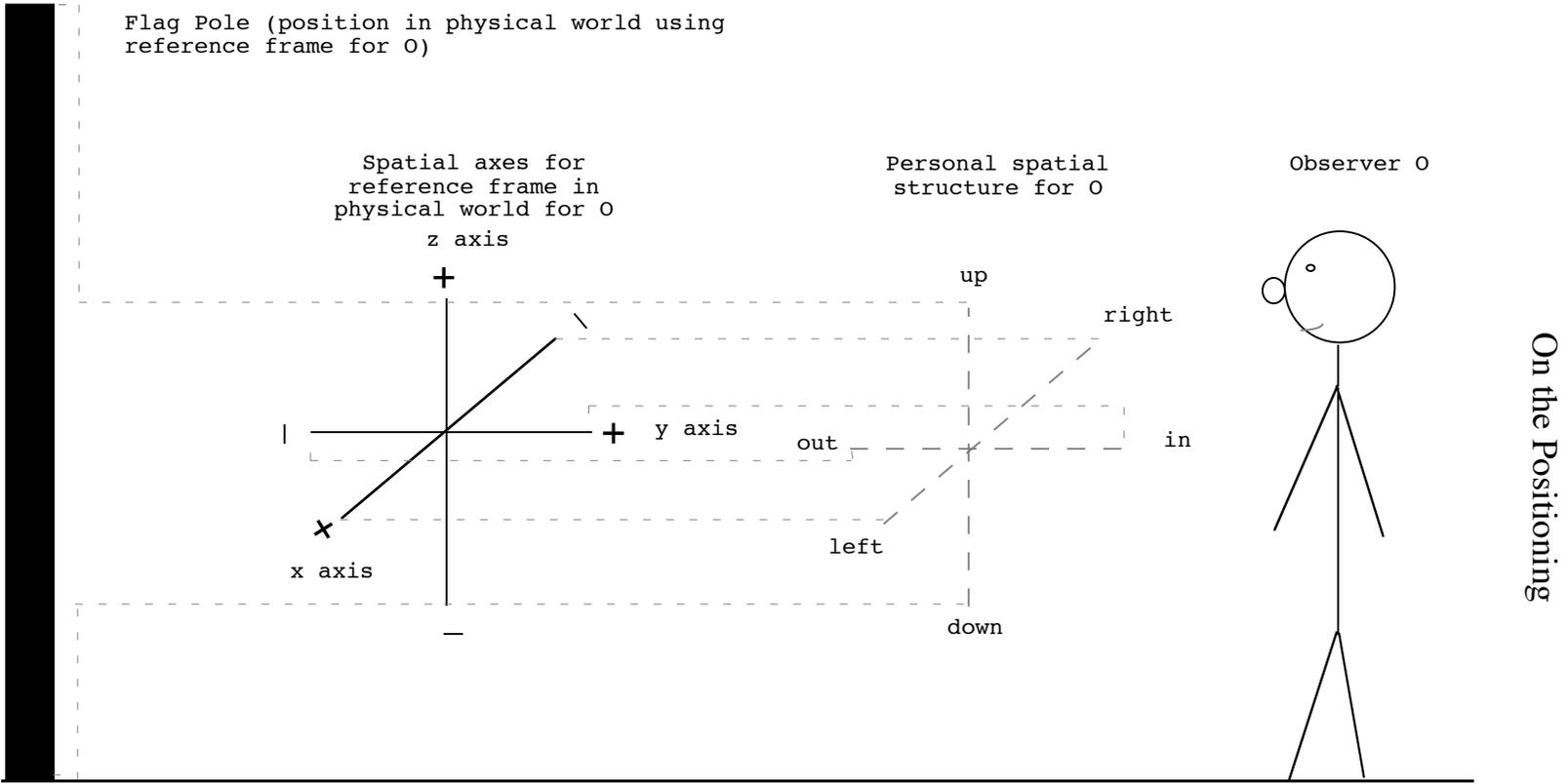

Figure 1. The correspondence of the $x$, $y$, and $z$ spatial
axes to the spatial dimensions height (i.e., up-down),
width (i.e., side-to-side), and depth (i.e., in-out) for observer O.

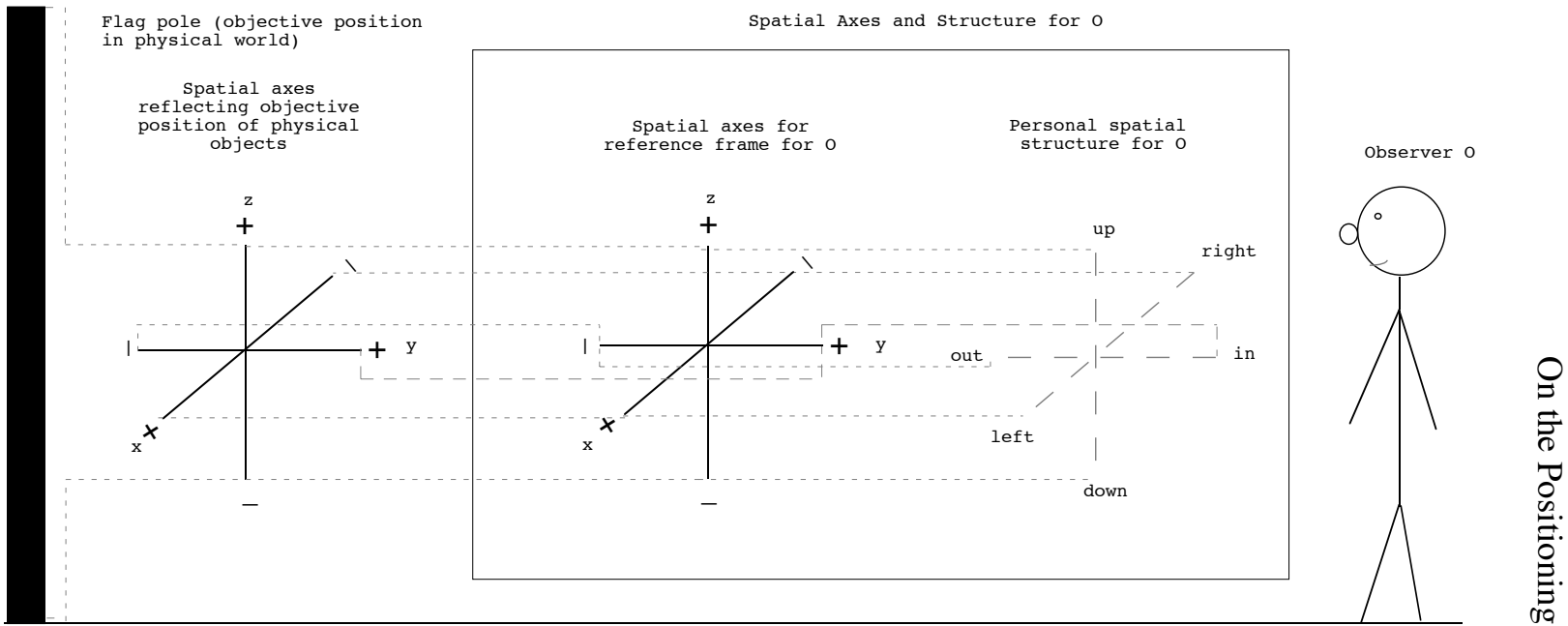

Figure 2.  The correspondence of the *x*, *y*, and *z*
spatial axes for observer O to the spatial axes *x*, *y*, and *z*
representing spatial structure in the physical world.

On the Positioning

# On the Positioning

Spacetime appears to the individual filtered through his personal spatial structure.[3,4]

Generally, physicists have not allowed that the spatial coordinate system in physics used to position a physical existent is influenced by the personal spatial structure of the individual. It is assumed that the structure of space, in particular its orientation, is unaffected by the perceptual and cognitive activities of the observer.[5] Figure 2 shows how the spatial structure of the individual might be accounted for if the structure of space is considered independent of the human observer and yet a "subjective" spatial structure and accompanying "subjective" spatial coordinate system is allowed.

---

[3] To get a sense of perception in other spatial reference frames, imagine a one dimensional spatial reference frame (i.e., a line). Let the line be straight and let the top of your extended index finger of your right hand represent this line. Place your extended index finger directly in front of your right eye while your left eye is shut. Raise the index finger so that you see the top of the index finger directly in front of you is a straight line. If you were a creature in this one dimensional reference frame represented by the top of your index finger, you would not see above or below the top surface of the finger. You would see in only one dimension. The metric (i.e., the unit length) before and behind you would be the same. As you travel on the line, you find the metric to be the same. Or if you move your finger in and out right next to your cheek, this accomplishes the same thing as traveling on the line. If the on the other hand, you bend your index finger like an inverted "u" and move your finger backwards and forwards (in and out) tracing the path of the inverted "u" in front of your eye, you also see just before you or behind you depending on whether you move your finger in or out. But the metric remains the same as you move your finger and you do not see off the top of the line. This is akin to traveling on the curved surface of your finger. Whether your finger is straight or curved, the metric is the same, and locally there is no difference between the straight and curved lines. This local metric being the same in all places results in calling the motion over the surface straight. It is straight for the surface. If one took a larger view, one can detect curvature within the reference frame of the curved finger. But this is a larger scale structure of the reference frame.

[4] It is possible that the personal spatial structure of an observer need not be Euclidean in nature, as phenomenological investigation has shown (e.g., Merleau-Ponty, 1945/1994). The implications of this finding on an observer's employment of spatial coordinate systems embodied in reference frames needs to be investigated.

[5] The alterations in the nature of space and time in relativity theory are generally considered to be dependent on factors other than human cognition. In relativity theory, though there is an acknowledgment of flexibility in the choice of reference systems, particularly in special relativity, this flexibility is generally limited to the choice of the reference frame, a spatiotemporal coordinate system, and is considered not to depend directly on an observer at rest in a reference frame. In the special theory, the relationship between inertial reference frames is generally considered to depend only on the invariant velocity of light in inertial reference frames in uniform translational velocity relative to one another without taking into account the reasoning necessary to account for the relativity of simultaneity (Snyder, 1994).

# On the Positioning

Our functioning in the physical world in our daily lives is remarkably good and dependent on our personal spatial structure being accurate in its application to the physical world.[6]  Where there is assumed to be an objective positioning of objects in space not influenced by the perceiving person, the correspondence of the "subjective" spatial structure to the "objective" positioning of objects in space would have to be good indeed.  The presumed dichotomy between the "objective" positioning of objects in space and a "subjective" spatial coordinate system is one basis for the dichotomy between a "subjective" spatiality and the objective space that many have maintained exists.[7]  Where a dichotomy is presumed, the "subjective" space may be in error, due to psychological factors, when it is considered in relationship to "objective" space.  In this view, one could conceivably have a situation like that depicted in Figure 3, showing the individual's "subjective" spatial perception to be in error.

Evidence previously cited supporting perceptual and behavioral adaptation of inversion of light to the eye supports the thesis that cognition affects the formal reference frames employed by individuals.  Other evidence concerns the phenomenon known as position constancy, that the perceived orientation of an object in space tends to remain the same despite different retinal images of the object (Rock, 1974/1997; Wallach, 1959).  But position constancy is affected by which directions the observer perceives up-down and right-left, for example.  An example is the perception of the form on the left in Figure 4 as a square where the square's top and bottom are extended along the horizontal of the manuscript page.  Referring to the "top" and "bottom" of the square indicates the positioning of the square in a personal spatial structure where, in terms of the square, up and down point in the directions shown in Figure 5.  If the square aligned along the vertical of the page is rotated 45° counterclockwise  (the right form in Figure 4)  and the two corners are perceived to point up and down, the figure is perceived to be a diamond.  Here up-down remains aligned along the length of the manuscript page, as shown in Figure 5.  If, on the other hand, the observer's sense of up-down and right-left has also rotated counterclockwise with  the form, the observer still

---

[6] It is not as accurate as various physical measuring systems are, but to a significant degree, people are able to gauge distance accurately.  This is what allows for our very precise motor behavior that is guided by our perception.

[7] This objective spatial coordinate system is not the absolute space of Newtonian mechanics. It is simply considered to be independent of cognition.

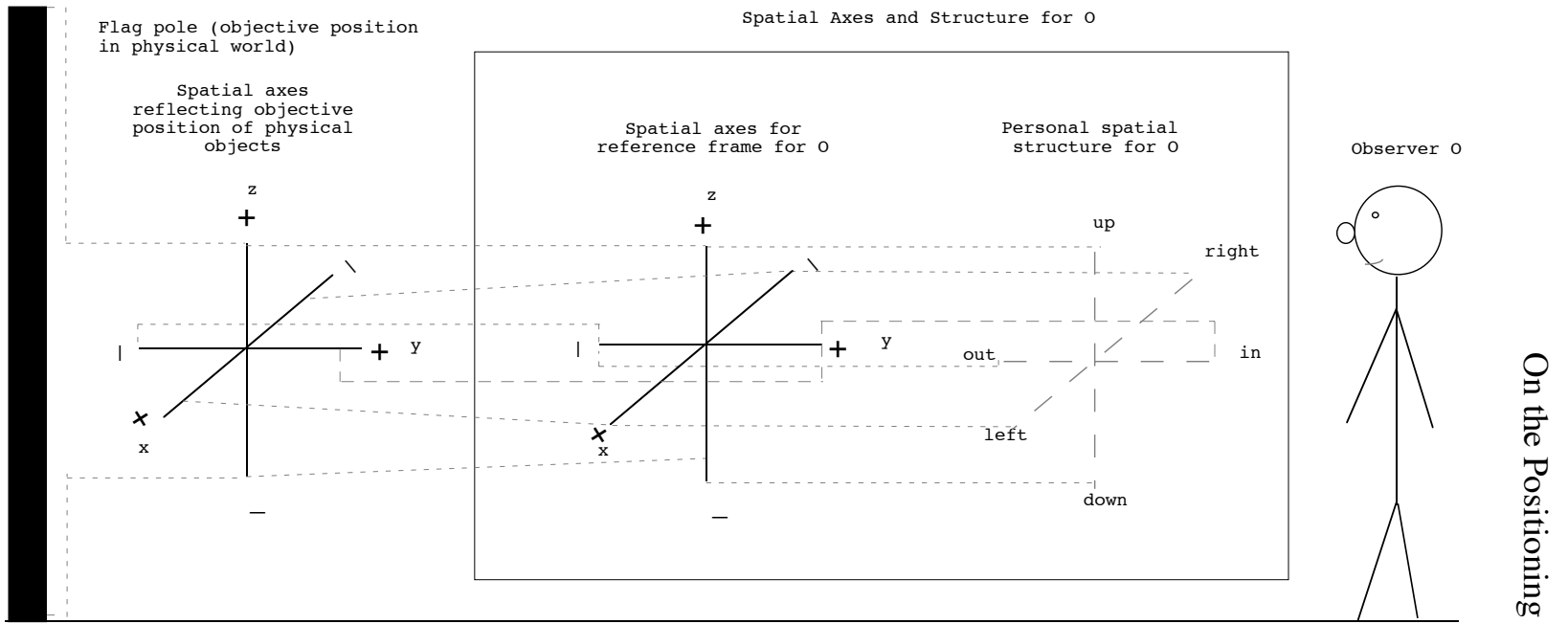

Figure 3. An example of "subjective" and "objective"
space represented by spatial coordinate systems
that do not maintain a uniform relationship.

perceives a square.  In the latter case, up and down point in the directions
indicated in Figure 6.  These directions (up-down and right-left) themselves
may be affected by other percepts that are themselves subject to factors
besides the retinal image.[8]

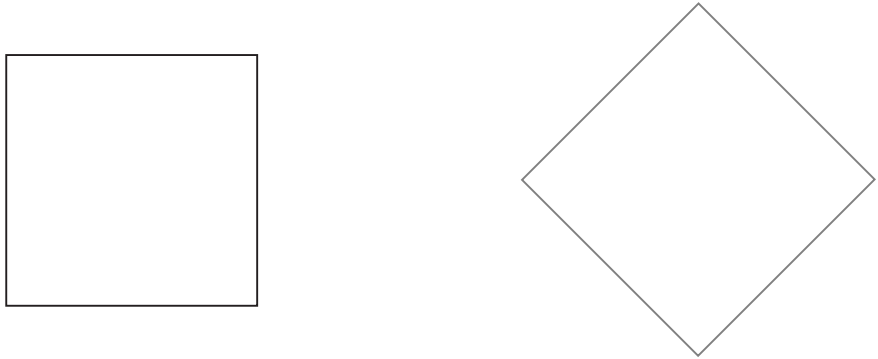

Figure 4. An example of
the relationship of form and
orientation.

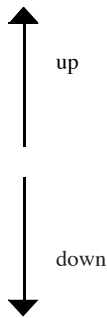

Figure 5. Up and down for the left form in Figure 4 and
where the right form is perceived as a diamond.

---

[8] Also, the example above indicates that spatial orientation may be layered.  That is,
perception of a figure depends on the spatial context relevant to that figure.

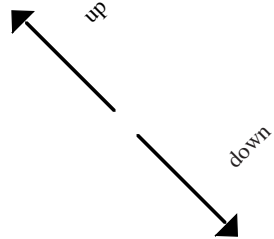

Figure 6. Up and down for right form in Figure
4 where it is perceived as a square.

In Figure 4, that the sides of the square on the left are equal in length is readily perceived due to the necessity of right angles in the square. For the figure on the right side, that the sides are of equal length is not so readily perceived. That they are equal comes after consideration of the figure, either of its angles, comparing the lengths of its sides, or seeing that the figure is a square when up and down are on the diagonal.

Other studies provide evidence of adaptation to alteration in incoming light and indicate the role of cognition in visual perception. For example, early on Helmholtz (1866/1925) reported a study in which a subject wore prisms that displaced objects laterally to the left from what would have been their normal position had the subject not worn the prisms. After repeated trials reaching for an object while looking at the object, the subject gained the ability to correctly touch the object with his eyes closed. Moreover, upon removal of the prisms, the subject had a negative aftereffect where the subject initially reached for the object in the direction too far to the right. After a certain number of failures, the subject adapted to not wearing the prisms.

In a series of studies involving tilting of objects in the environment relative to the subject (Asch & Witkin, 1948a, 1948b; Witkin & Asch, 1948a, 1948b), the roles of orientation of the visual field relative to the observer, as well as gravity, in perception of the upright have been demonstrated, indicating an integrative feature of perception. Similarly, the perception of motion depends on whether an object's perceived change of location can be accounted for by head and/or eye movement. Again, changing the position of the retinal image does not in and of itself account for motion, as an observer's tracing a moving object keeps the retinal image of the object essentially in the same retinal location. Rock (1974/1997) discussed the square and diamond percepts noted above in the context of a theory of "indirect" perception. In a number of experiments performed by Rock and others, it has been shown that perception is a cognitive endeavor, where percept-to-percept relationships form a hierarchical backbone to that which an individual finally perceives (Rock, 1997).

Though it has been demonstrated that the personal spatial structure and reference frame used by an individual are affected by cognition, what does this have to do with the measurement of space and time in physics? The key is that in experiments like those conducted by Stratton, where *all incoming light* is rotated by some constant degree, the observer is in a new situation when adaptation occurs.[9] *Adaptation implies that there is no objective spatial structure independent of the observer.* If there were an objective spatial structure and given adaptation, how would an observer know it?[10] It was because of the apparent difficulty posed by this question that psychologists maintained at the end of the last century that the visual system was hardwired. Then, perceptual spatial orientation (linked to only one form of neurophysiological activity) would be absolute and correlate with objective

---

[9] In Stratton's experiments, light was inverted, that is rotated 180° by the use of a lens system.

[10] Consider the following observation reported by the subject in Snyder and Pronko's study, who happened to be Snyder, that supports adaptation. This observation also indicates that something has changed for the subject with inversion of light, but only when the historical event of the inversion of light to the subject is attended to.

> Toward the end of the experiment [i.e., the period in which the subject wore the inverting glasses], the subject was adequately adjusted [adapted]. The following insightful experience occurred. He was observing the scene from a tall building. Suddenly someone asked, "Well, how do things look to you? Are they upside-down?"
>
> The subject replied, "I wish you hadn't asked me. Things were all right until you popped the question at me. Now, when I recall how they *did* look *before* I put on these lenses, I must answer that they do look upside down now. But until the moment that you asked me I was absolutely unaware of it and hadn't given a thought to the question of whether things were right-side-up or upside-down." (Snyder & Pronko, 1952, p. 113)

An historical event, namely the donning of inverting glasses, presents the basis for knowing that there is a *difference* between normal presentation of incoming light and inversion of incoming light. This event has meaning for the subject, Snyder, in the above quote through his *memory* of the pre-inversion situation. This memory does not affect the subject's perception and perceptually-guided action where the historical event of inverting incoming light is *not* attended to.

space in the physical world. This would ensure our ability to accurately gauge the spatial relations inherent in the physical world.

Demonstrating that perceptual constancy, including what a figure is, can be altered by spatial rotation of the object, and can be restored by a corresponding rotation of the observer's sense of up-down and right-left, provides support that a figure's spatial orientation (whether it is up, down, or in some other orientation) and attendant effects regarding figural spatial extension for example are affected by the observer. This is a key finding in supporting perceptual and behavioral adaptation to inversion of incoming light. First, this result supports the possibility that the spatial orientation of an observer's entire visual world, and the changes in spatial structure that result, can change. Second, as these effects are due to the observer, the possibility of adaptation and its being accounted for in terms of a shift of the observer's personal spatial structure is supported. Because of the significance of the observer in affecting spatial structure, the thesis that the formal spatial structure represented in the frame of reference he employs is associated directly with the adapted observer's personal spatial structure is supported. The distinction between an "objective" spatial structure and a "subjective" one is not supported. The use of a reference frame in making precise measurements reflects the actual position of objects in the physical world for that observer. This situation is like that portrayed in Figure 1.

One might still ask, if everything comes out okay in the end, that is everything is still lined up correctly in the personal spatial structure, the individual's "subjective" reference frame, and an "objective" spatial structure in the physical world, then of what significance is this finding? First, the finding brings the human observer into line with the implications of the special theory of relativity on the significant role of the human observer in the development of spacetime structure (Snyder, 1994).[11] Second, the finding indicates that a primary result in the theory of quantum mechanics is limited in scope. That is, with this finding, it becomes possible to know two quantities that within quantum mechanics are mutually exclusive, albeit in two different formal reference frames reflecting different personal spatial structures. For example, the spin components for an electron along orthogonal spatial axes may be simultaneously known, albeit within two different formal reference frames (Snyder, 1992, 1993, 1995). This finding definitely holds for different observers. Because of the possibility of biperceptual capabilities concerning simultaneous personal spatial structures for an individual (e.g., Dolezal, 1982), this finding allows for the possibility that a single observer may know both quantities through employing two distinct reference frames simultaneously. Investigating psychological variables concerning perception can lead to understanding the physical world itself in addition to perception.[12]

---

[11] As noted, that the human observer is significant in the development of spacetime curvature is not the generally accepted view.

---

[12] The finding that cognition affects the structure of space and time (or spacetime) is discussed in more detail in *The Mind and the Physical World: A Psychologist's Exploration of Modern Physical Theory* (Snyder, 1996) and *An Essay on the Relationship between the Mind and the Physical World* (Snyder, 1997).